\documentclass[pre,twocolumn,amsmath,amssymb]{revtex4}

\usepackage{graphicx}

\newcommand{\lang}{\left\langle}
\newcommand{\rang}{\right\rangle}

\begin{document}

\title{Comment on ``On the Crooks fluctuation theorem and the Jarzynski equality'' [J. Chem. Phys. 129, 091101 (2008)]}

\author{Artur B. Adib}
\email{adiba@mail.nih.gov}
\affiliation{
Laboratory of Chemical Physics, NIDDK, National Institutes of Health, Bethesda, Maryland 20892-0520, USA
}

\date{\today}

\begin{abstract}
It has recently been argued that a self-consistency condition involving the Jarzynski equality (JE) and the Crooks fluctuation theorem (CFT) is violated for a simple Brownian process [L. Y. Chen, {\em J. Chem. Phys.} {\bf 129}, 091101 (2008)]. This note adopts the definitions in the original formulation of the JE and CFT and demonstrates the contrary.
\end{abstract}

\maketitle

Consider the problem of an overdamped Brownian particle moving under the action of a time-dependent potential $U(x,t)$ over the whole line $-\infty < x < \infty$ in the time interval $0 \leq t \leq \tau$. Let us define the following objects of interest (for simplicity of notation, I will set $k_B T=1$):

(a) The {\em free energy difference}
\begin{equation} \label{def_deltaF}
  \Delta F := - \ln \frac{Z_\tau}{Z_0},
\end{equation}
where the {\em partition function} $Z_t$ corresponding to the instantaneous energy surface $U(x,t)$ is defined as
\begin{equation}
  Z_t := \int_{-\infty}^{\infty} \! dx \, e^{-U(x,t)};
\end{equation}

(b) The {\em work functional} defined along a given trajectory $x(t)$,
\begin{equation} \label{def_W}
  W[x(t)] := \int_0^\tau \! dt \, \frac{\partial U(x(t),t)}{\partial t},
\end{equation}
where the partial derivative $\partial/\partial t$ keeps $x(t)$ constant; and lastly

(c) The {\em Jarzynski average}, defined as the average $\langle \cdot \rangle$ over all possible trajectories $x(t)$ propagated under $U(x,t)$ from $t=0$ to $t=\tau$, where $x(0)$ is sampled from the initial Boltzmann distribution $p(x(0)) = e^{-U(x(0),0)}/Z_0$. (An expression for such an average in terms of e.g. path integrals can be immediately written down, but this is not relevant for the present discussion).

Making explicit and unambiguous use of definitions (a)-(c), the Jarzynski equality (JE) \cite{jarzynski97-prl}
\begin{equation} \label{jarz}
  \lang e^{-W} \rang = e^{-\Delta F},
\end{equation}
has been proven through several different approaches in the literature (see \cite{jarzynski08-epjb} for a review).

Previous derivations notwithstanding, a skeptic might question the validity of Eq.~(\ref{jarz}) at two rather distinct levels; namely, the appropriateness of the definitions (a)-(c), or---once such definitions are agreed upon---the correctness of the ensuing equality itself. The first is mostly a matter of historical context and semantics, and will not be the subject of the present paper (see e.g. \cite{vilar08-prl,jarzynski08-comment} for a discussion on definitions (a)-(b)). The second, on the other hand, would imply some type of mathematical inconsistency in the steps leading up to Eq.~(\ref{jarz}).

In Ref.~\cite{chen08a}, the two levels of inquiry above have been blurred together, leading its author to---among other misconceptions---claim that Eq.~(\ref{jarz}) is ``violated for a simple model system driven far from equilibrium.'' What is missing in such claim is the necessary qualification that the original underpinnings of the equality, i.e. definitions (a)-(c), have been tampered with. The object of this note is to point out that precisely the opposite conclusion is reached if one uses the original definitions.

To proceed, consider the {\em forward} and {\em reverse energy functions}, $U_f(x,t)$ and $U_r(x,t)$ respectively, satisfying the condition
\begin{equation} \label{Ufr}
  U_f(x,t) = U_r(x,\tau - t),
\end{equation}
for $0\leq t \leq \tau$. This mapping between energy functions is the continuum analog of Crooks's forward and reverse processes \cite{crooks98}, and was presumably implicit in the treatment of Ref.~\cite{chen08a}. (When such energy functions are used instead of $U(x,t)$ in the definitions (a)-(b), the appropriate subscripts $f$ and $r$ will be added to the corresponding quantities; for (c), i.e. the average $\lang \cdot \rang$, the energy function is specified by the subscript of the quantity inside the brackets).

Since the forward and reverse energy functions coincide at the opposite ends of the time interval, i.e. $U_f(x,0)=U_r(x,\tau)$ and $U_f(x,\tau)=U_r(x,0)$, we have trivially $\Delta F_r = -\Delta F_f$, and thus the JE leads directly to Eq.~(4) of Ref.~\cite{chen08a}:
\begin{equation} \label{chen1}
  \lang e^{-W_f} \rang\, \lang e^{-W_r} \rang = 1.
\end{equation}
It is worth emphasizing that this result exists independently of the Crooks fluctuation theorem (CFT) or its generalizations, such as Eq.~(1) of Ref.~\cite{chen08a}; indeed, as we have just seen, it is merely an application of the Jarzynski equality twice for energy functions that coincide at opposite ends of the time interval, as expressed above. Nonetheless, for problems satisfying Eq.~(\ref{Ufr}) and microscopic reversibility, its breakdown would imply that both the CFT and the JE are violated, as for such problems the former also requires Eq.~(\ref{chen1}) to be true \cite{chen08a}. 

The central exercise of Ref.~\cite{chen08a} was to consider a simple choice of $U(x,t)$ that allows for analytical computations, and test the validity of the JE and CFT through Eq.~(\ref{chen1}). This program was carried out indirectly by considering systems whose work distributions are known to be Gaussian, in which case the exponential averages of work in Eq.~(\ref{chen1}) can be straightforwardly reduced to their first and second moments. This Gaussian assumption leads finally to Chen's self-consistency condition (Eq.~(6) of Ref.~\cite{chen08a}), namely
\begin{equation} \label{chen2}
  \lang W_f \rang + \lang W_r \rang = \frac{1}{2}\left( \sigma_f^2 + \sigma_r^2 \right),
\end{equation}
where $\sigma_f^2 \equiv \langle (W_f - \langle W_f \rangle)^2 \rangle$ and similarly for $\sigma_r^2$. In the following, I will consider the same model system of Ref.~\cite{chen08a}, and compute such moments analytically using the original definitions (a)-(c).

Chen's system can be described by the forward energy function
\begin{equation} \label{Uf}
  U_f(x,t) = \frac{k}{2}x^2 - f_0 \theta(t) \, x.
\end{equation}
This is a simple harmonic potential whose center $z(t) = f_0 \theta(t) / k$ is instantaneously displaced, at $t=0$, from the initial position $z(0) = 0$ to the final position $z(\tau)=f_0/k$. [I am using the convention that the step function $\theta(t)$ vanishes at $t=0$, which is consistent with Chen's choice of initial conditions, viz. $\langle x(0) \rangle = 0$. This convention implies, in particular, that $\int_0^\tau \! dt \, x(t) \dot{\theta}(t)=x(0)$ for any finite $\tau >0$, a result that is used in Eqs.~(\ref{Wf}) and (\ref{Wr})]. Accordingly, $U_f(x,0) = \frac{k}{2}x^2$, and with definition (b) we have
\begin{equation} \label{Wf}
  W_f[x(t)] = -f_0 \int_0^\tau \! dt \, x(t) \delta(t) = -f_0 \, x(0).
\end{equation}
Thus, the moments of $W_f$ reduce to moments of the initial coordinate $x(0)$, which give
\begin{equation} \label{results1}
  \lang W_f \rang = 0, \quad \lang \sigma_f^2 \rang = \frac{f_0^2}{k}.
\end{equation}
Similarly, with the reverse energy function (cf. Eq.~(\ref{Ufr}))
\begin{equation}
  U_r(x,t) = \frac{k}{2}x^2 - f_0 \theta(\tau-t) \, x,
\end{equation}
we have
\begin{equation} \label{Wr}
  W_r[x(t)] = f_0 \int_0^\tau \! dt \, x(t) \delta(\tau-t) = f_0 \, x(\tau).
\end{equation}
Since the original equilibrium state specified by $U_r(x(0),0)$ is not perturbed until $t=\tau$, the coordinate $x(\tau)$ is distributed like $x(0)$, and thus
\begin{equation} \label{results2}
  \lang W_r \rang = \frac{f_0^2}{k}, \quad \lang \sigma_r^2 \rang = \frac{f_0^2}{k}.
\end{equation}
With the results given by Eqs.~(\ref{results1}) and (\ref{results2}), Chen's self-consistency condition (Eq.~(\ref{chen2})) is immediately satisfied, q.e.d.

Some final remarks are in order. Though never unambiguously stated in Ref.~\cite{chen08a}, it seems like the analysis offered by Chen departed from the original formulation of the Jarzynski equality and the Crooks fluctuation theorem in the definition of work, i.e. in definition (b) of the present note. Using the original definition, the free energy difference $\Delta F$ obtained via the Jarzynski equality and Eqs.~(\ref{results1}) and (\ref{results2}) (cf. Eq.~(5) of Ref.~\cite{chen08a}) is $-f_0^2/(2k)$ for the forward, and $+f_0^2/(2k)$ for the reverse process, in agreement with the different approach of Ref.~\cite{jarzynski08-comment}, and consistent with the simple observation that the free energy difference between two states changes sign upon a change of direction in the process. Had Chen adhered to the original definitions, no controversy would have arisen.

The author would like to thank Attila Szabo for bringing Chen's manuscript to his attention. This research was supported by the Intramural Research Program of the NIH, NIDDK.

\end{document}